3

\def\D{D\!\!\!\!/}

\def\P{\partial\!\!\!/}
\documentstyle[12pt,a4]{article}
\begin{document}

\begin{center}
{\Large\bf CHIRAL ANOMALIES IN FIELD THEORIES\footnote{Invited article
for a book to be published by the Indian National Science Academy
on the occasion of the International Mathematical Year 2000.}
} \\
\vspace{1cm}
{\bf H. Banerjee} \\
{S. N. Bose National Centre for Basic Sciences} \\
{Salt Lake, Calcutta - 700 091, India} \\
{e-mail : banerjee@boson.bose.res.in}
\end{center}
\vspace{2cm}

\begin{abstract}
The role of the contribution from the fermion mass term in the
axial vector Ward identity in generating the U(1) axial anomaly,
both local and global, is elucidated. Gauge invariance requires
the fermion to decouple from the gauge field if it is very heavy.
This identifies the Adler-Bell-Jackiw (ABJ) anomaly with the
asymptotic limit of the sign reversed mass term. In an instanton
background, the chiral limit $(m = 0)$ of the mass term does not
vanish but consists of contributions from fermion zero modes.
Space time integral of these zero mode contributions exactly
cancels, thanks to the Atiyah-Singer index theorem, the integral
of the ABJ anomaly and suggests that the Jacobian for  global
U(1) chiral transformation is trivial even in an instanton
background. This can be realised in the representation of the
fermion partition function in a Weyl basis. The resolution of the
strong CP problem is thus achieved in an axionless physical
world.

In chiral gauge theories the fermion partition function admits of
a gauge invariant representation but only at the cost of locality.
Implementation of fermion averaging of the gauge current with the
invariant partition function yields the current whose covariant
derivative is the covariant anomaly. With the covariant current
as input one can derive an integrable current whose covariant
derivative is the minimal consistent anomaly obeying the
Wess-Zumino consistency condition. The distinction between the
two currents disappears if either the covariant or the consistent
anomaly vanishes. This is realised only if the fermion belongs to
an anomaly-free representation of the gauge group.
\end{abstract}

\newpage
\baselineskip=20pt
\section{Introduction}
In classical field theories there is a correspondence between a
global symmetry of the action and a conserved Noether current.
Presence of short distance singularities which need to be
regularised for mathematical consistency complicates matters in
quantum field theory (QFT). It may so happen that a
regularisation scheme with mandatory attributes, like gauge
invariance in a gauge theory of fermions, and at the same time
consistent with the global symmetry cannot be formulated or
simply does not exist. Traces of violation of the global symmetry in the form of non-conservation of the Noether current may
survive as the regulator is removed at the end of calculation.
This is the genesis of anomalies and anomalous Ward identities in
QFT.

The topic of anomaly, in particular, axial anomaly came on the
centrestage of particle physics research through the studies of
neutral pion decay into two photons. The decay rate $1.2 \times
10^{16}$ per sec. was explained satisfactorily by
Steinberger$^{1}$ in 1949 in terms of triangle diagrams (Fig.1)
with proton circulating in the fermion loop. The linear
divergence of the amplitude was regulated by the Pauli--Villars
method. Problem arose sixteen years later$^{2}$, when decay rates
obtained within the framework of current algebra and partial
conservation of axial vector current (PCAC) were invariably
smaller than the data by three orders of magnitude.

A popular working hypothesis, PCAC derives its dynamical basis in
gauge theories of fermion like quantum chromodynamics (QCD) from
the `naive' operator relation (or, equivalently, naive Ward
identity)
$$\partial_{\mu}
\left(\bar{q}\gamma_{5}\gamma_{\mu}\tau_{3}q\right) =
2m\left(\bar{q}\gamma_{5}\tau_{3}q\right) \eqno\ldots(1.1)$$
which follows from field equations, with $q$ the quark doublet $(u,d)$
and $\tau_{3}$ the isospin. One recognises in the left hand side
the Noether current corresponding to the chiral symmetry $u
\rightarrow e^{i\alpha\gamma_{5}}u, d \rightarrow
e^{-i\alpha\gamma_{5}}d$, which should be conserved at the
classical level in the chiral limit $m = 0$ of QCD. PCAC is just
the statement that the mass term on the right hand side of (1.1)
can be replaced by the neutral pion field
$$\partial_{\mu}\left(\bar{q}\gamma_{5}\gamma_{\mu}\tau_{3}q\right)
= F_{\pi}m^{2}_{\pi}\pi^{0} \eqno\ldots(1.2)$$
where $F_{\pi}$ is the pion decay constant, $m_{\pi}$ the pion
mass, and $\pi^{0}$ the pion field. This is an unexceptionable
step since the mass term has the right quantum numbers of a
neutral pion and, therefore, can be regarded as the definition of
the pion field in terms of quark constituents.

Problem with PCAC in $\pi^{0} \rightarrow 2\gamma$ stemmed from
the Sutherland--Veltman$^{3}$ theorem which states that
substitution of the divergence of isospin axial current for the
neutral pion field in the matrix element yields a null result for
the decay rate. Coupled with the positive result of
Steinberger$^{1}$, the unambiguous conclusion that emerges from the
theorem is that the inadequacy of the PCAC relation stems really
from the naive relation (1.1) which is flawed if quarks participate in
electromagnetic interactions. The missing element was diagnosed
as an anomaly, the Adler-Bell-Jackiw (ABJ) anomaly$^{4}$ in the
Noether current for chiral symmetry
$$\partial_{\mu}\left(\bar{q}\gamma_{5}\gamma_{\mu}\tau_{3}q\right)
= 2m\left(\bar{q}\gamma_{5}\tau_{3}q\right) -
\left(\frac{N_{c}}{3}\right)\frac{e^{2}}{16\pi^{2}}\epsilon_{\mu\nu\alpha\beta}
F_{\mu\nu} F_{\alpha\beta} \eqno\ldots(1.3)$$
where $N_{c}$ is the colour degree of freedom of quarks and
$F_{\mu\nu}$ the electromagnetic field tensor. The ABJ anomaly,
therefore, modifies the `naive' PCAC relation (1.2) to
$$F_{\pi}m^{2}_{\pi}\pi^{0} =
\partial_{\mu}\left(\bar{q}\gamma_{5}\gamma_{\mu}\tau_{3}q\right)
+ \left(\frac{N_{c}}{3}\right)\frac{e^{2}}{16\pi^{2}}
\epsilon_{\mu\nu\alpha\beta} F_{\mu\nu} F_{\alpha\beta}$$
The decay rate now calculated by substituting the anomaly term
for the pion field in the matrix element for $\pi^{0} \rightarrow
2\gamma$ is given by
$$\Gamma\left(\pi^{0} \rightarrow 2\gamma\right) =
\left(\frac{N_{c}}{3}\right)^{2} \times 1.11 \times 10^{16}
\sec^{-1} \eqno\ldots(1.4)$$
Depending on how one looks at the result (1.4), it may be
regarded as either a signal success of the diagnosis of the problem in
$\pi^{0} \rightarrow 2\gamma$ as due to anomaly, or in the light
of later developments, a prediction of the number of colour
degrees of freedom $N_{c} = 3$ in QCD.

Success in $\pi^{0} \rightarrow 2\gamma$ problem brought into
limelight the scenario of breaking symmetries at the classical
level through anomalies in quantum field theories. Gauge
theories become inconsistent if gauge symmetry is violated
through anomaly. Cancellation of anomalies, therefore,
constitutes an important constraint in building models for
physical gauge theories with chiral coupling to fermions. Global
chiral anomaly seems to play a key role in discussing physical
effects associated with topologically nontrivial gauge
field configurations.

\section{Axial Anomaly and Fermion Decoupling}
In a gauge theory of fermion there is a contradiction at the
quantum level between chiral invariance and gauge symmetry. The
ABJ anomaly, or in the more general context of non-Abelian gauge
theories of fermion, the anomaly in the U(1) axial vector
current, arises because gauge invariance is to be preserved for
consistency of the theory. The contradiction is transparent in
the condition for decoupling$^{5}$ of the fermion from the
background gauge field when it is very heavy. For the divergence
of the U(1) axial vector current the decoupling condition assumes
the form of an anomalous Ward identity
$$\langle\partial_{\lambda}\left(\bar{\psi}(x)\gamma_{5}\gamma_{\lambda}
\psi(x)\right)\rangle = 2m \langle\bar{\psi}(x)\gamma_{5}\psi(x)\rangle -
\lim_{m\rightarrow\infty}\left[2m\langle\bar{\psi}(x)\gamma_{5}\psi(x)\rangle\right]
\eqno\ldots(2.1)$$
where $\langle$ $\rangle$ denotes that the fermion degrees of
freedom are integrated out. As we shall see below, (2.1) follows
directly from gauge invariance and is known as Adler$^{6}$
condition in QED. Eq.(2.1) will still be compatible with chiral
symmetry and a conserved U(1) axial vector current would emerge
in the chiral limit $m = 0$ if the second term on the right
hand side were to vanish. But this is not to be. The asymptotics
in field theory gives in the infinite mass limit the ABJ anomaly
$$\lim_{m\rightarrow\infty}\left[2m\langle\bar{\psi}(x)\gamma_{5}\psi(x)\rangle\right]
= \frac{g^{2}}{16\pi^{2}} \epsilon_{\mu\nu\lambda\rho} tr
F_{\mu\nu} F_{\lambda\rho} \eqno\ldots(2.2)$$
where $F_{\mu\nu} = F_{\mu\nu}^{a} t_{a}$ are the field tensors
with $t_{a}$ the generators of the gauge group.

\begin{figure} 
\unitlength1cm
\begin{minipage}[t]{8cm}
\begin{picture}(10,8)
\put(-0,-20){\includegraphics{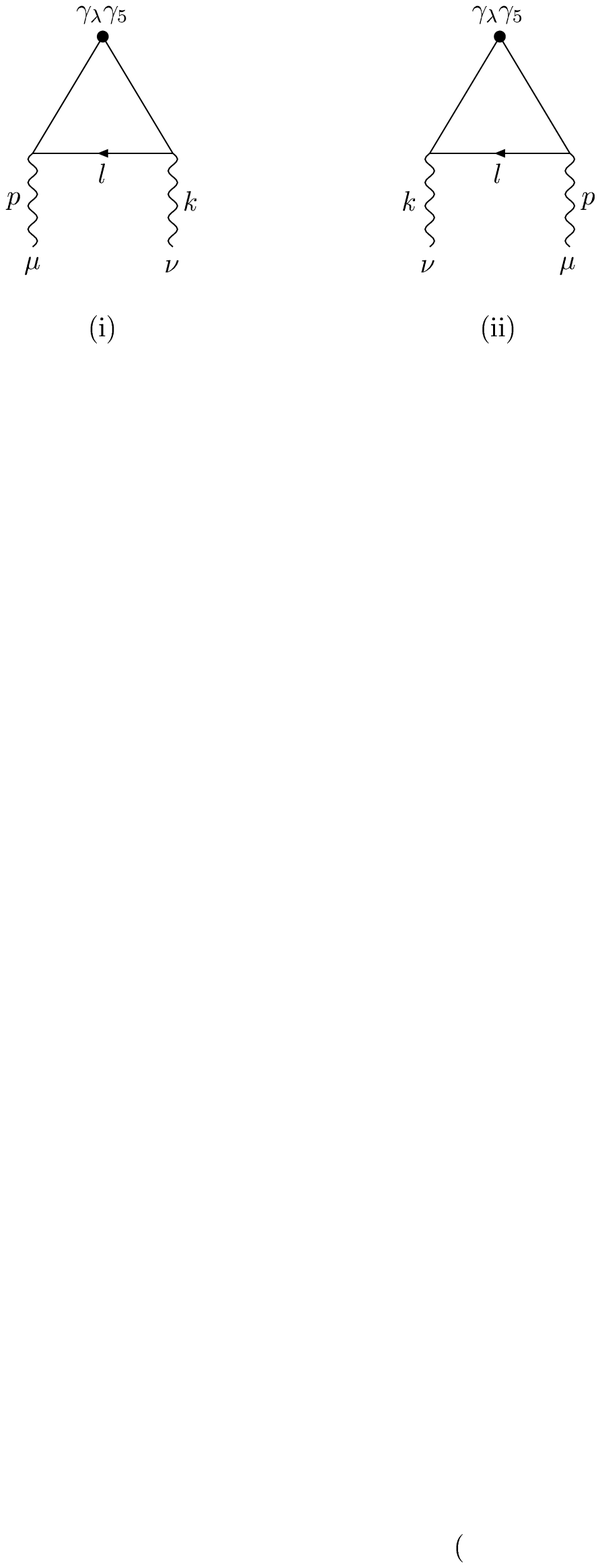}}
\end{picture}
\end{minipage}
\caption{Triangle diagrams }
\end{figure}


To motivate the decoupling condition (2.1) we consider in QED the
amplitude for creating two photons with momenta and polarisation
$(p, \mu)$ and $(k,\nu)$ by the axial vector current ${J_{\lambda
5}(x)=\bar{\psi}(x)\gamma_{5}\gamma_{\lambda}\psi(x)}$ 
$$\langle p,\mu; k,\nu \vert J_{\lambda 5}(0)\vert 0 \rangle =
\epsilon_{\mu}(p)\epsilon_{\nu}(k)M_{\lambda\mu\nu}(p,k,m)$$
The key to the analysis is the Rosenberg$^{7}$ tensor
decomposition (see Fig.1)
$$M_{\lambda\mu\nu} = \epsilon_{\lambda\mu\nu\alpha} k_{\alpha}
A(p,k,m) + \epsilon_{\lambda\nu\alpha\beta} p_{\alpha} k_{\beta}
\left[p_{\mu} B(p,k,m) + k_{\mu}C(p,k,m)\right] + \left[(k,\nu)
\leftrightarrow (p,\mu)\right] \eqno\ldots(2.3)$$
which follows from parity and Lorentz invariance.

Not all the form factors are independent. The form factor $A$
which gives the divergence of the axial vector current
$$(p + k)_{\lambda} M_{\lambda\mu\nu} =
\epsilon_{\mu\nu\alpha\beta} p_{\alpha}k_{\beta} \left[A(p,k,m)
+ A(k,p,m)\right] \eqno\ldots(2.4)$$
is determined through gauge invariance by the form factors $B$
and $C$
$$A(p,k,m) = p^{2}B(p,k,m) + p.k C(p,k,m) \eqno\ldots(2.5)$$

The form factors $B$ and $C$ are of dimensions [mass]$^{-2}$, and,
therefore, in perturbation theory they are represented by highly
convergent amplitudes which vanish as $m^{-2}$ for large fermion
mass
$$\lim_{m\rightarrow\infty} B(p,k,m) = \lim_{m\rightarrow\infty}
C(p,k,m) = 0$$
Thus gauge invariance (2.5) guarantees that the divergence of the
amplitude for the axial vector current given in (2.4) vanishes
in the asymptotic $m\rightarrow\infty$ limit
$$\lim_{m\rightarrow\infty} (p + k)_{\lambda} M_{\lambda\mu\nu} =
0 \eqno\ldots(2.6)$$

In perturbation theory the amplitude $M_{\lambda\mu\nu}$ for the
triangle diagram is linearly divergent. The leading divergence,
however, drops out due to symmetric integration of loop momentum
leaving a potential logarithmic divergence, which can appear only
in the form factor $A$ in (2.3). Gauge invariance (2.5) rules out
even this residual logarithmic divergence.

The above observations suggest that the potential anomaly
represented by the second term in (2.1) must be finite and
independent of regularisation scheme. To verify this we start by
calculating
$$\lim_{m\rightarrow\infty}\left[2m\langle\bar{\psi}(x)\gamma_{5}\gamma(x)\rangle\right]
= \lim_{m\rightarrow\infty}\left[2m\langle x\vert Tr \gamma_{5}
(i\D + m)^{-1} \vert x\rangle\right], \eqno\ldots(2.7)$$

where, to conform to our discussions in the subsequent sections,
we work in Euclidean metric and write for the hermitian Dirac
operator
$$\D =\gamma_{\mu}\left(i\partial_{\mu} - gA_{\mu}\right)
\eqno\ldots(2.8)$$
with $A_{\mu} \equiv A^{a}_{\mu}t_{a}$, the gauge potential.

Our strategy is to develop the Green function appearing in (2.7)
in a perturbative series
$$(i\D + m)^{-1} = (-i\D + m) G$$
with
$$\begin{array}{lll}
G &= &\left(\D^{2} + m^{2}\right)^{-1} \\
&= &G_{0} - gG_{0}VG_{0} + g^{2}G_{0}VG_{0}VG_{0} + \ldots
\end{array} \eqno\ldots(2.9)$$
The `free' part is $G_{0} = \left(p^{2} + m^{2}\right)^{-1}$ with
$p_{\mu} = i\partial_{\mu}$ the `momentum' operator. The
`potential' $gV$ has two pieces
$$gV = gV_{0} + \frac{1}{2} \sigma_{\mu\nu} F_{\mu\nu}$$
with $\sigma_{\mu\nu} =
\frac{1}{2i}\left(\gamma_{\mu}\gamma_{\nu} -
\gamma_{\nu}\gamma_{\mu}\right)$. The first piece $gV_{0}$ is at
most linear in $p$ and independent of $\gamma$--matrices. The
trace with $\gamma_{5}$ in (2.7) starts to be nonvanishing only
from terms of order $g^{2}$ onwards in the perturbative expansion
(2.9) and one obtains (2.2)
$$\lim_{m\rightarrow\infty}
\left[2m\langle\bar{\psi}(x)\gamma_{5}\psi(x)\rangle\right] =
\lim_{m\rightarrow\infty}\left[2m^{2}\langle x\vert
Tr\left(\gamma_{5}G\right)\vert x\rangle\right] =
\frac{g^{2}}{16\pi^{2}} \epsilon_{\mu\nu\lambda\rho} tr
F_{\mu\nu}(x) F_{\lambda\rho}(x)$$
Note that the final result is local. All nonlocalities as well as
contributions from higher order terms in the perturbative series
(2.9) drop out in the infinite mass limit $m\rightarrow\infty$.

In the decoupling condition (2.1) one can set the mass term on
the right hand side to zero in the chiral limit $m=0$. The
anomalous Ward identity thus obtained
$$\langle\partial_{\lambda}\left(\bar{\psi}\gamma_{5}
\gamma_{\lambda}\psi\right)\rangle_{m=0}
= -\frac{g^{2}}{16\pi^{2}}\epsilon_{\mu\nu\lambda\rho} tr
F_{\mu\nu} F_{\lambda\rho} \eqno\ldots(2.10)$$
shows that the U(1) axial vector current, i.e. the Noether current
corresponding to global chiral symmetry
$$\psi \rightarrow e^{i\alpha\gamma_{5}}\psi, \quad \bar{\psi}
\rightarrow \bar{\psi}e^{i\alpha\gamma_{5}}, \eqno\ldots(2.11)$$
of the massless Dirac operator (2.8), is not conserved. The
divergence of the current is just the ABJ anomaly which is
responsible for the two photon decay of neutral pion discussed in
the preceding section.

We note that in renormalisable theories in perturbative framework
the decoupling condition (2.1), which is a special example of the
decoupling theorem of Appelquist and Carazzone$^{5}$, is correct
to all orders of perturbation, just as the Adler--Bardeen$^{8}$
theorem assures us that the anomalous axial vector Ward identity
(2.10) is not affected by radiative corrections in QED.

It should be remarked that setting the mass term to zero in
(2.1), as was done in obtaining (2.10), may not always be
legitimate in the chiral limit $m=0$ if the gauge field is
treated nonperturbatively. The Euclidean Dirac operator (2.8) has
zero modes if the background gauge field has a nontrivial
topology. In this scenario the chiral limit of the mass term does
not vanish and, as we shall see in the next section, consists
precisely of the zero modes of the Dirac operator.

\section{Path Integral Approach to Anomaly}
In a seminal work Fujikawa$^{9}$ interpreted the ABJ anomaly
within the path integral framework as arising from the nontrivial
Jacobian of the fermion measure under chiral transformation. In
Euclidean metric, considered by Fujikawa, the partition function
which generates fermion Green functions in a background gauge
field configuration may be written as
$$Z_{f}[A] \equiv \int d\mu \exp \left[\int\bar{\psi}(i\D
+ m)\psi d^{4}x\right] \eqno\ldots(3.1)$$
where $d\mu$ is the integration measure for fermion and
$\D$ is the Euclidean Dirac operator defined in (2.8).
Fermion Green functions which are normalised expectation values
of any product $O$ of fermion fields are obtained from the
functional integral (3.1)
$$\langle O \rangle = \frac{1}{Z_{f}[A]} \int d\mu O \exp
\left[\int\bar{\psi}(i\D + m)\psi d^{4}x\right]
\eqno\ldots(3.2)$$
With $\gamma$-matrices chosen hermitian, the Dirac operator (2.8)
is also hermitian with real eigenvalues $\lambda_{n}$ and
orthonormal eigenfunctions $\phi_{n}(x)$
$$\D\phi_{n}(x) = \lambda_{n}\phi_{n}(x),
\quad \int\phi^{+}_{m}(x) \phi_{n}(x) d^{4}x = \delta_{mn}
\eqno\ldots(3.3)$$
Each nonzero eigenvalue $\lambda_{n}$ has its chirally conjugate
partner $-\lambda_{n}$ with eigenfunctions $\phi_{-n}$
$$\D\phi_{-n} = -\lambda_{n}\phi_{-n}, \quad \phi_{-n} =
\gamma_{5}\phi_{n} \eqno\ldots(3.4)$$
In perturbative field theories one is interested in gauge field
configurations with only trivial topology. For such
configurations the kernel space of the Euclidean Dirac operator
(2.8) is of dimension zero. This means that the set $\{\phi_{n}(x)\}$
with nonzero eigenvalues constitute a complete basis in
function space. The Dirac field $\psi(x)$ can be expanded in this
basis as
$$\psi(x) = \sum\left(a_{n} + a_{-n}\gamma_{5}\right)\phi_{n}(x)
\eqno\ldots(3.5)$$
where $a_{\pm n}$ are complex-valued Grassmann generators. The
four degrees of freedom corresponding to each mode of the Dirac
field is accounted for if we split $a_{\pm n}$ as
$$a_{\pm n} = \alpha_{\pm n} + i\beta_{\pm n}$$
with $\alpha, \beta$ real valued.

There are ambiguities$^{10, 11, 12}$ on the issue whether in
Euclidean metric $\bar{\psi}$ should be treated as independent of
$\psi$. For the present, we follow the popular ansatz$^{10}$ and
expand $\bar{\psi}(x)$ with an independent set of Grassmann
generators $\{\bar{b}_{\pm n}\}$,
$$\bar{\psi}(x) = \sum \phi^{+}_{n}(x) \left(\bar{b}_{n} +
\bar{b}_{-n}\gamma_{5}\right) \eqno\ldots(3.6)$$
The representations for $\psi(x)$ and $\bar{\psi}(x)$ together
with orthonormality of the eigenfunctions (3.3) yield for the
fermion action
$$\begin{array}{ll}
S_{f}(m) &= \int\bar{\psi}(x) (i\D + m)\psi(x) d^{4}x \\
&= \sum\left[\left(i\lambda_{n} + m\right)\bar{b}_{n} a_{n} +
\left(-i\lambda_{n} + m\right)\bar{b}_{-n} a_{-n}\right]
\end{array} \eqno\ldots(3.7)$$
The integration measure for the fermion fields in the basis
$\{\phi_{n}(x)\}$ is
$$d\mu \equiv \Pi_{n} d\bar{b}_{n}da_{n}d\bar{b}_{-n}da_{-n}
\eqno\ldots(3.8)$$
The standard rules of integration of Grassmann generators now
yield for the partition function (3.1) the desired result
$$\begin{array}{ll}
Z_{f}[A] &= \Pi_{n}\left(\lambda^{2}_{n} + m^{2}\right) \\
&= \det\left(i\D + m\right) \end{array} \eqno\ldots(3.9)$$
This confirms the correctness of the choice of the measure (3.8).

Ward identities, whether normal or anomalous, are obtained in
path integral framework from the requirement that the partition
function is invariant under infinitesimal symmetry
transformations of the variables of integration. To derive the
Ward identity corresponding to chiral symmetry one implements a
`local' chiral transformation of the variables of integration
$\psi(x), \bar{\psi}(x)$ in the partition function (3.1)
$$\begin{array}{ll}
\psi(x) \rightarrow \psi'(x) &= \left(1 +
i\alpha(x)\gamma_{5}\right)\psi(x) \\
\bar{\psi}(x) \rightarrow \bar{\psi}'(x) &= \bar{\psi}(x) \left(1
+ i\alpha(x)\gamma_{5}\right) \end{array} \eqno\ldots(3.10)$$
The fermion measure (3.8) changes and the new measure
corresponding to the transformed variables of integration is
given by $d\mu'$
$$\begin{array}{ll}
d\mu' &= \Pi_{n}d\bar{b}'_{n} da'_{n} d\bar{b}'_{-n} da'_{-n} \\
&= d\mu \hspace{.2cm} J[\alpha], \end{array} \eqno\ldots(3.11)$$
where $a'_{n}(\bar{b}'_{n})$ are the new set of Grassmann
generators in the expansion of $\psi'(x)
\left(\bar{\psi}'(x)\right)$ in the basis $\{\phi_{n}(x)\}$. The
Jacobian $J[\alpha]$ can be calculated  following standard procedure
$$J[\alpha] = \exp\left[- 2 i\int d^{4}x
\alpha(x)\sum_{n}\left(\phi^{+}_{n}(x)\gamma_{5}\phi_{n}(x) +
\phi^{+}_{-n}(x)\gamma_{5}\phi_{-n}(x)\right)\right]
\eqno\ldots(3.12)$$
The fermion action also changes and the new action is given by
$$S_{f}(m) \rightarrow S'_{f}(m) = \int d^{4}x
\left[\bar{\psi}(i\D + m)\psi + i\alpha(x)
\left(-\partial_{\mu}\left(\bar{\psi}\gamma_{5}\gamma_{\mu}\psi\right)
+ 2m\bar{\psi}\gamma_{5}\psi\right)\right] \eqno\ldots(3.13)$$
Invariance of the partition function (3.1) under the
infinitesimal local chiral transformation (3.10) now gives the
anomalous axial Ward identity
$$\langle\partial_{\mu}\left(\bar{\psi}\gamma_{5}\gamma_{\mu}\psi\right)\rangle
= 2m \langle\bar{\psi}\gamma_{5}\psi\rangle -2\sum_{n}
\left(\phi^{+}_{n}\gamma_{5}\phi_{n} +
\phi^{+}_{-n}\gamma_{5}\phi_{n}\right) \eqno\ldots(3.14)$$
It is easy to recognise (3.14) as the decoupling condition (2.1)
of the preceding section. Indeed
$$\begin{array}{ll}\displaystyle
\lim_{m\rightarrow\infty}
\left[2m\langle\bar{\psi}\gamma_{5}\psi\rangle\right] &=
\displaystyle \lim_{m\rightarrow\infty}
\left[2m\sum_{n}\left(\frac{\phi^{+}_{n}\gamma_{5}\phi_{n}}{m   
+ i\lambda_{n}} + \frac{\phi^{+}_{-n}\gamma_{5}\phi_{-n}}{m -
i\lambda_{n}}\right)\right] \\
&= 2\sum_{n}\left(\phi_{n}^{+}\gamma_{5}\phi_{n} +
\phi^{+}_{-n}\gamma_{5}\phi_{-n}\right) \end{array}
\eqno\ldots(3.15)$$
where the infinite sum on the right hand side of (3.14) is to be cut off
gauge invariantly, $\vert\lambda_{n}\vert \ll M$ for $M$ large.
Fujikawa used the gauge invariant cut off
$\exp\left(-\D^{2}/M^{2}\right)$ with large $M$ to evaluate
the infinite sum
$$\begin{array}{ll}
2\sum_{n}\left(\phi^{+}_{n}(x)\gamma_{5}\phi_{n}(x) +
\phi^{+}_{-n}(x)\gamma_{5}\phi_{-n}(x)\right) &=
\displaystyle\lim_{M\rightarrow\infty} 2\langle x\vert
\gamma_{5}e^{-\D^{2}/M^{2}} \vert x\rangle \\
&= \displaystyle\frac{g^{2}}{16\pi^{2}}
\epsilon_{\mu\nu\lambda\rho} tr F_{\mu\nu}(x) F_{\lambda\rho}(x)
\end{array} \eqno\ldots(3.16)$$ 
It is clear that the left hand side of (3.16) should be augmented
by zero modes if the Dirac operator admits of them. Zero modes
always appear with definite chiralities, $\epsilon_{i} = \pm 1$
$$\D \hspace{.2cm} \phi_{0i} = 0, \quad \gamma_{5}\phi_{0i} =
\epsilon_{i}\phi_{0i} \eqno\ldots(3.17)$$
This is because, in its kernel space the Dirac operator commutes
with $\gamma_{5}$. In the presence of zero modes the left hand
side of (3.16) needs to be augmented by their contributions,
i.e.,
$$\frac{g^{2}}{16\pi^{2}} \epsilon_{\mu\nu\lambda\rho} tr
F_{\mu\nu}(x) F_{\lambda\rho}(x) =
2\left[\sum_{i}\epsilon_{i}\phi^{+}_{0i}\phi_{0i} + \sum_{n}
\left(\phi^{+}_{n}\gamma_{5}\phi_{n} +
\phi^{+}_{-n}\gamma_{5}\phi_{-n}\right)\right]
\eqno\ldots(3.18)$$
Space time integral of (3.18) gives the Atiyah--Singer$^{13}$
index theorem
$$\nu \equiv \frac{g^{2}}{32\pi^{2}} \epsilon_{\mu\nu\lambda\rho} \int
d^{4}x tr F_{\mu\nu}(x) F_{\lambda\rho}(x) = n_{+} - n_{-}
\eqno\ldots(3.19)$$
where $\nu$ is the winding number (Pontryagin index) of the gauge
field and $n_{+} (n_{-})$ is the number of positive (negative)
chirality zero modes. Eigenmodes corresponding to nonzero eigenvalues
do not contribute to the space time integral (3.19)
because $\phi_{n}$ is orthogonal to $\gamma_{5}\phi_{n}$. Note
that nontrivial winding number, $\nu \not= 0$, is realised
through instanton--like configuration of the gauge field.

The presence of zero modes has profound impact on the chiral
limit of the fermion mass term on the right hand side of the anomalous
axial Ward identity
$$\langle\partial_{\mu}(\bar{\psi}\gamma_{5}\gamma_{5}\psi)\rangle
= 2m\langle\bar{\psi}\gamma_{5}\psi\rangle -
\frac{g^{2}}{16\pi^{2}} \epsilon_{\mu\nu\lambda\rho} tr
F_{\mu\nu} F_{\lambda\rho} \eqno\ldots(3.20)$$
The zero modes can be isolated from the mass term
$$2m\langle\bar{\psi}\gamma_{5}\psi\rangle = 2m
\langle\bar{\psi}\gamma_{5}\psi\rangle' + 2\sum
\epsilon_{i}\phi^{+}_{0i}\phi_{0i} \eqno\ldots(3.21)$$
where the first term on the right hand side is bereft of the
zero modes and vanishes in the chiral limit
$$2m\langle\bar{\psi}\gamma_{5}\psi\rangle' = 4\sum_{\lambda_{n}
> 0} \frac{m^{2}}{m^{2} + \lambda_{n}^{2}} \phi^{+}_{n}
\gamma_{5}\phi_{n} \eqno\ldots(3.22)$$
The mass term, therefore, has now a nontrivial chiral limit
consisting precisely of the zero modes, and the chiral limit of
the axial Ward identity is not exactly what was obtained in the
perturbative framework of the preceding section
$$\langle\partial_{\mu}
(\bar{\psi}\gamma_{5}\gamma_{\mu}\psi)\rangle_{m=0} = 2\sum
\epsilon_{i}\phi^{+}_{0i}\phi_{0i} - \frac{g^{2}}{16\pi^{2}}
\epsilon_{\mu\nu\lambda\rho} tr F_{\mu\nu} F_{\lambda\rho}
\eqno\ldots(3.23)$$
The zero modes in the extra piece appearing on the right hand
side arise from instanton--like configuration of the gauge field
and, therefore, could not have been accessed in a perturbative
framework.

It is of great interest to note that the zero mode terms which
appear explicitly on the right hand side of (3.23) are exactly
cancelled by similar terms contained now (see (3.18)) in the ABJ
anomaly. Thus, irrespective of whether or not the gauge field
configuration gives rise to zero modes, the chiral limit of the
local axial anomaly comprises of only nonzero eigenmodes of the
Dirac operator
$$\langle\partial_{\mu}(\bar{\psi}\gamma_{5}\gamma_{\mu}\psi)\rangle_{m=0}
= -2\sum_{\vert\lambda_{n}\vert\not= 0}
\left(\phi^{+}_{n}\gamma_{5}\phi_{n} +
\gamma^{+}_{-n}\gamma_{5}\phi_{-n}\right) \eqno\ldots(3.24)$$
We, therefore, conclude that the space-time integral of the
chiral limit of the divergence of the axial vector current always
vanishes. This follows from the orthogonality of $\phi_{n}$ and
$\gamma_{5}\phi_{n}$ if one uses (3.24) or from the
Atiyah--Singer index theorem if instead one uses (3.23)
$$\int\langle\partial_{\mu}(\bar{\psi}\gamma_{5}\gamma_{5}\psi)\rangle_{m=0}
d^{4}x = 0 \eqno\ldots(3.25)$$
This, as we shall see later, has a profound impact on issues of
physics related to global chiral anomaly.

\section{Chiral Gauge Theories and the Covariant and Consistent
Anomalies}
The ABJ anomaly in the U(1) axial vector Ward identity
constitutes an unambiguous evidence of a fundamental
incompatibility of chiral invariance and gauge symmetry in
regularisation scheme in perturbative framework of quantum field
theory. In a vector-like gauge theory, such as QCD, chiral
invariance is an expendable attribute and the ABJ anomaly results
from strict adherence to gauge symmetry. In chiral gauge theories
where gauge fields are coupled chirally to fermions in the Dirac
operator
$$\begin{array}{ll}
{\cal D} &\equiv \gamma_{\mu}\left(i\partial_{\mu} + t_{a}
A^{a}_{\mu} \frac{1}{2} (1 - \gamma_{5})\right) \\
&= \left(i\P + /\!\!\!\! A \frac{1}{2} (1  - \gamma_{5})\right), \end{array}
\eqno\ldots(4.1)$$
loss of chiral invariance jeopardises gauge symmetry and hence
the consistency of the theory. The fermion action
$$S_{F} = \int\bar{\psi}{\cal D}\psi d^{4}x \eqno\ldots(4.2)$$
is invariant under the local chiral gauge transformations
$$\displaystyle \psi(x) \rightarrow e^{i\alpha(x)\frac{(1 -
\gamma_{5})}{2}} 
\psi(x), \quad \bar{\psi}(x) \rightarrow \bar{\psi}(x)
e^{-i\alpha(x)\frac{(1 + \gamma_{5})}{2}}$$
$$A_{\mu}(x) \rightarrow e^{i\alpha(x)} \left\{A_{\mu}(x) +
\frac{1}{i}\partial_{\mu}\right\}e^{-i\alpha(x)}
\eqno\ldots(4.3)$$
with $\alpha(x) = t_{a}\alpha_{a}(x)$ the gauge function.
Dimensional regularisation, popular in perturbative gauge
theories, has serious problem with $\gamma_{5} =
\frac{1}{4!}\epsilon_{\mu\nu\lambda\rho}
\gamma_{\mu}\gamma_{\nu}\gamma_{\lambda}\gamma_{\rho}$. The
totally antisymmetric tensor of rank four
$\epsilon_{\mu\nu\lambda\rho}$ does not admit of suitable
generalisation to arbitrary space-time dimensions. Thus, one is
yet to find a consistent and systematic scheme for regulating
divergences in chiral gauge theories in weak coupling
perturbation in the continuum.

On lattice, the finite spacing $a$ between lattice
sites provides a built-in regularisation of all short distance
singularities in field theories. Here too, the prospects for a
consistent formulation of chiral gauge theory are not really
bright. The major problem on lattice is the species doublers of
fermion and their removal. The doublers appear as unwanted zeros
of the Fourier transform of the `free' Dirac operator on lattice,
over and above the zero at the origin of momentum space which
correspond to the physical fermion. In the `naive' Dirac operator
$\left(\gamma_{\mu}\sin\left(p_{\mu}a\right)/a\right)$ the
doublers are located at the edges of the Brillouin zone
\hspace{.2cm} $-(\pi/a)
\leq p_{\mu} \leq (\pi/a)$. \hspace{.2cm} The doublers are not specific for the
naive Dirac operator. According to the celebrated theorem of
Nielsen and Ninomiya$^{14}$ these are generic and can be avoided only
at a price, by breaking explicitly locality and/or chiral
symmetry in the Dirac operator. The most popular model for
lattice fermion, the Wilson model$^{15}$, removes the doublers by
giving them masses of the order of the lattice cut-off $O(1/a)$
$$D_{W}(p) = \gamma_{\mu}\sin(p_{\mu}a)/a + ir\left(1 -
\cos(p_{\mu}a)\right)/a \eqno\ldots(4.3)$$
Gauge invariance is implemented simply through link variables as
in all lattice models. But the explicit breaking of chiral
symmetry for nonzero `$r$' makes the model patently inappropriate
for chiral gauge theories. Current spurt in interest in the
subject stems mainly from the realisation  that for lattice Dirac
operators $D$ obeying the Ginsparg--Wilson$^{16}$ relation
$$\gamma_{5}D + D\gamma_{5} = aD\gamma_{5}D, \eqno\ldots(4.4)$$
chiral symmetry is restored and species doublers are removed
in the continuum limit$^{17}$. The issue of nonlocality implied in
the Ginsparg--Wilson relation, particularly in the context of
chiral gauge theories, is yet to be resolved $^{18}$.

{\bf Covariant Anomaly}$^{19}$ : Apart from the absence of a
consistent and systematic regularisation scheme, chiral gauge
theories are, in general, afflicted with anomalies in the gauge
current. The Dirac operator (4.1) in chiral gauge theory is
non-hermitian. A fallout of this is that Fujikawa's$^{9}$ recipe
for constructing a gauge invariant partition function, which
assumes a hermitian Dirac operator, needs to be modified. The
Dirac operator ${\cal D}$ in (4.1) maps $\psi$ into the space of
spinors in the domain of ${\cal D}^{+}$. The eigenvalue
equations (3.3) are, therefore replaced by
$${\cal D}\phi_{n} = \lambda_{n}\chi_{n}, \quad \quad {\cal
D}^{+}\chi_{n} = \lambda_{n}\phi_{n}, \eqno\ldots(4.5)$$
where $\lambda^{2}_{n}$ are real, nonnegative and constitute the
eigenvalue spectrum of ${\cal D D}^{+}$ and ${\cal
D}^{+}{\cal D}$. The sets of eigen functions $\{\phi_{n}\}$
and $\{\chi_{n}\}$ of ${\cal D}^{+}{\cal D}$ and ${\cal D
D}^{+}$ respectively constitute an orthonormal basis for
expanding $\psi$ and $\bar{\psi}$
$$\psi = \sum a_{n}\phi_{n}, \quad \bar{\psi} =
\sum_{n}\bar{b}_{n}\chi^{+}_{n} \eqno\ldots(4.6)$$
in terms of the Grassmann generators $a_{n}, \bar{b}_{n}$. The
fermion measure defined as
$$d\mu[A] = \Pi_{n} d\bar{b}_{n} da_{n} \eqno\ldots(4.7)$$
is a gauge invariant functional of $A_{\mu}$ and yields the
partition function$^{19}$
$$\begin{array}{ll}
Z_{inv}[A] &\equiv \int d\mu[A] \exp [\int\bar{\psi}{\cal D}\psi
d^{4}x] \\
&= \left(\det{\cal D}^{+}{\cal D}\right)^{1/2} = \left(\det
{\cal D D}^{+}\right)^{1/2} \end{array} \eqno\ldots(4.8).$$
Both ${\cal D}^{+}{\cal D}$ and ${\cal D D}^{+}$ change by
a similarity transformation under gauge transformation. The
representation (4.8) is thus formally  gauge invariant.

The chiral gauge current
$$\left[\bar{\psi}t_{a}\gamma_{\mu}\frac{1}{2}(1 -
\gamma_{5})\psi\right] \eqno\ldots(4.9)$$
transforms covariantly under gauge transformation (4.3). Fermion
averaging of the current with the gauge invariant measure (4.7)
yields
$$\begin{array}{ll}
J^{a}_{\mu}(x) &\equiv \displaystyle\frac{\int d\mu[A]
\left(\bar{\psi}t_{a}\gamma_{\mu}\frac{1}{2}(1 -
\gamma_{5})\psi\right)\exp\left[\int\bar{\psi}{\cal D}\psi
d^{4}x\right]}{\int d\mu[A]\exp[\int\
\bar{\psi}{\cal D}\psi
d^{4}x]} \\
&= \displaystyle\sum_{n} \frac{1}{\lambda_{n}}
\chi^{+}_{n}t_{a}\gamma_{\mu}\frac{1}{2}(1 - \gamma_{5})\phi_{n}.
\end{array} \eqno\ldots(4.10)$$
Gauge invariant regularisation can be implemented by suppressing
large eigenvalues and the current thus obtained transforms
covariantly and is called the covariant current.

Formal application of field equations suggest that the gauge
current should be covariantly conserved. This, however, may not
be true for the fermion averaged current $J^{a}_{\mu}(x)$ if it
is anomalous,
$$\begin{array}{ll}
G^{a}(x) &\equiv \partial_{\mu} J^{a}_{\mu}(x) -
f^{abc}A^{b}_{\mu}(x)J^{c}_{\mu}(x) \\
&= \sum_{n}\left\{\chi^{+}_{n}t_{a}\frac{1}{2}(1 +
\gamma_{5})\chi_{n} - \phi^{+}_{n}t_{a}\frac{1}{2}(1 -
\gamma_{5})\phi_{n}\right\} \end{array} \eqno\ldots(4.11)$$
Following Fujikawa's$^{9}$ recipe for gauge invariant
regularisation one obtains the covariant anomaly
$$\begin{array}{ll}
G^{a}(x) &= \displaystyle\lim_{M\rightarrow \infty}
\int\frac{d^{4}k}{(2\pi)^{4}} Tr t_{a} \left[\frac{1}{2}(1 +
\gamma_{5}) e^{ik.x} e^{-\frac{{\cal D}{\cal D}^{+}}{M^{2}}} e^{-ik.x}
\frac{1}{2}(1 + \gamma_{5}) e^{ik.x} e^{-\frac{{\cal
D}^{+}{\cal D}}{M^{2}}} e^{-ik.x}\right] \\
&= \displaystyle-\frac{1}{32\pi^{2}} \epsilon_{\mu\nu\lambda\rho}
tr [t_{a} 
F_{\mu\nu} F_{\lambda\rho}] \end{array} \eqno\ldots(4.12)$$
where $F_{\mu\nu} = t_{a}F^{a}_{\mu\nu}$ are the field tensors.

{\bf Consistent Anomaly}$^{19,20}$: In perturbative treatment of
chiral gauge theories the fermion measure in the partition
function is independent of the gauge field. A fallout of this is
that, unlike $Z_{inv}[A]$ in (4.8), the perturbative partition
function
$$\begin{array}{ll}
Z_{pert}[A] &\equiv e^{W[A]} \\
&= \int d\mu \exp \left[\int\bar{\psi}{\cal D}\psi d^{4}x\right],
\end{array} \eqno\ldots(4.13)$$
and hence the effective action $W[A]$ need not be gauge invariant. The
gauge current with fermion averaging implemented through this
perturbative partition function
$$\begin{array}{ll}
J^{a}_{W\mu}(x) &\equiv \displaystyle\frac{\delta}{\delta
A^{a}_{\mu}(x)} 
W[A]. \\
&= \langle\bar{\psi} t_{a}\gamma_{\mu}\frac{1}{2}(1 -
\gamma_{5})\psi\rangle_{W} \end{array} \eqno\ldots(4.14)$$
will, in general, not transform covariantly. However, it must
obey the integrability condition
$$\frac{\delta J^{a}_{W\mu}(x)}{\delta A^{b}_{\nu}(x')} -
\frac{\delta J^{b}_{W\nu}(x')}{\delta A^{a}_{\mu}(x)} = 0,
\eqno\ldots(4.15)$$
since it is defined in (4.14) through the functional derivative
of the effective action $W[A]$. The current $J^{a}_{W\mu}(x)$ is
called the consistent current and its covariant derivative
$$G^{a}(x) \equiv \partial_{\mu}J^{a}_{W\mu}(x) -
f^{abc}A^{b}_{\mu}J^{c}_{W\mu}(x) \eqno\ldots(4.16)$$
is the consistent anomaly.

Gauge transformation properties of an arbitrary functional of
gauge fields are best discussed with the help of the generators
$$L^{a}(x) = \partial_{\mu}\frac{\delta}{\delta A^{a}_{\mu}(x)} -
f^{abc} A^{b}_{\mu}(x) \frac{\delta}{\delta A^{c}_{\mu}(x)}
\eqno\ldots(4.17)$$
Thus the consistent anomaly $G^{a}_{W}(x)$, representing as it
does the gauge variation of the effective action $W[A]$, is given
by
$$G^{a}_{W}(x) = L^{a}(x) W[A] \eqno\ldots(4.18)$$
The algebra of the generators
$$\left[L^{a}(x), L^{b}(x')\right] = f^{abc}\delta^{a}(x - x') L^{c}(x)
\eqno\ldots(4.19)$$
shows that the consistent anomaly must obey the Wess-Zumino$^{21}$
consistency condition
$$L^{a}(x) G^{b}_{W}(x') - L^{b}(x')G^{a}_{W}(x) =
f^{abc}\delta^{4}(x - x')G^{c}_{W}(x) \eqno\ldots(4.20)$$
On the other hand, the anomaly $G^{a}_{W}(x)$ is a measure of the
non-covariance of the consistent current $J^{a}_{W\mu}(x)$
$$L^{b}(x') J^{a}_{\mu}(x) = -f^{abc}\delta^{4}(x - x') +
\frac{\delta G^{b}_{W}(x')}{\delta A^{a}_{\mu}(x)}
\eqno\ldots(4.21)$$

As for the covariant anomaly (4.12), one finds, as expected, an
incompatibility with the Wess-Zumino consistency condition
$$L^{a}(x) G^{b}(x') - L^{b}(x')G^{a}(x) = 2f^{abc}\delta^{4}(x -
x')G^{c}(x), \eqno\ldots(4.22)$$
where the factor 2 on the right hand side spoils consistency.
Thus, the anomaly itself is a measure of the `inconsistency'. The
origin of the `inconsistency' may be traced to the fermion
measure $d\mu[A]$ given by (4.7) for averaging of the gauge
current in the definition (4.10) of the covariant current
$J^{a}_{\mu}(x)$. A nontrivial covariant anomaly $G^{a}(x)$
corresponds to a nontrivial dependence of the measure $d\mu[A]$
on the gauge field. This is suggested also from the observation
that the definition
$${\cal J}^{a}_{\mu}(x) \equiv \frac{\delta}{\delta A^{a}_{\mu}(x)} ln
Z_{inv}[A] \eqno\ldots(4.23)$$
where $Z_{inv}[A]$ is the gauge invariant partition function
(4.8), has all the attributes, it is covariant, consistent and
anomaly free. The price that one pays for this `perfect' current
is a high degree of nonlinearity.

It can be shown$^{19}$ that the consistent current coincides with
the covariant current if the functional curl of the latter
vanishes
$$J^{a}_{W\mu}(x) = J^{a}_{\mu}(x) + \int^{1}_{0}dg \int d^{4}x'
A^{b}_{\nu}(x') \left\{\frac{\delta J_{\nu}^{bg}(x')}{\delta
A^{a}_{\mu}(x)} - \frac{\delta J^{ag}_{\mu}(x)}{\delta
A^{b}_{\nu}(x')}\right\} \eqno\ldots(4.24)$$
where $J^{ag}_{\mu}(x)$ is the covariant current corresponding to
the Dirac operator ${\cal D}^{g} = \left(i\P +
g/\!\!\!\!A \frac{1}{2}(1 - \gamma_{5})\right)$ with coupling
constant $g$. One can obtain from (4.24) an explicit
representation of the consistent anomaly using the expression
(4.12) for the covariant anomaly$^{19}$
$$G^{a}_{W}(x) = f^{1}_{0}dg G^{ag} +
\frac{1}{16\pi^{2}}\epsilon_{\mu\nu\lambda\rho} \int^{1}_{0}dg
g(1-g) tr \left(\left[t_{a},
A_{\mu}\right]\left(F^{g}_{\lambda\rho} A_{\nu} +
A_{\nu}F^{g}_{\lambda\rho}\right)\right) \eqno\ldots(4.25)$$

The above analysis shows that the distinction between covariant
and consistent currents disappears if and only if the anomaly in
either current vanishes. The fundamental requirement that the chiral
gauge theory is free of either anomaly imposes the unique
constraint on the group generators of the chiral fermions
$$tr\left(t_{a} \left\{t_{b}, t_{c}\right\}\right) = 0
\eqno\ldots(4.26)$$
which is symmetric in all the indices. An interesting application
in the Standard Model is to take $t_{a} = Q$, the matrix of
electric charge, and $t_{b}, t_{c}$ the isospin matrices. The
constraint $tr Q = 0$ is obeyed in the Standard Model since
each generation of quark doublet of three colours is paired with
a lepton doublet.

\section{Global Chiral Anomaly and the Strong CP Problem.}

Global U(1) axial anomaly is the sine qua non for the strong CP
problem. The problem consists in the gross disagreement in the
experimental data for the CP violating electric dipole moment of
neutron (EDMN) which are consistent with a null result and
theoretical estimates that invariably give a large value. Strong
CP problem provides the unique arena where the concept of a
global chiral anomaly is confronted with direct experimental
data.

The two possible sources for CP violation in QCD action
$$S_{QCD} = S_{G} + \int\bar{q}(i\D)qd^{4}x + m\int\bar{q}e^{2
i\alpha_{ew}\gamma_{5}}q d^{4}x + \theta_{QCD}\Delta S
\eqno\ldots(5.1)$$
are the chiral phase $\alpha_{ew}$ in the quark mass which
arises from the electroweak sector of the Standard Model, and the
QCD vacuum term with parameter $\theta_{QCD}$
$$\theta_{QCD}\Delta S = \theta_{QCD}\frac{g^{2}}{32\pi^{2}}
\epsilon_{\mu\nu\lambda\rho} \int tr
F_{\mu\nu}F_{\lambda\rho}d^{4}x \eqno\ldots(5.2)$$
In (5.1) $S_{G}$ represents the contributions from the gauge
fields. For gauge fields with nontrivial topology the coefficient
of $\theta_{QCD}$ in (5.2) gives precisely the winding number
$\nu\not= 0$,
$$\nu = \frac{g^{2}}{32\pi^{2}} \epsilon_{\mu\nu\lambda\rho} \int
tr F_{\mu\nu}F_{\lambda\rho}d^{4}x \eqno\ldots(5.3)$$
The chiral phase in the mass term in (5.1) can be transformed
away by relabelling the quark fields
$$q \rightarrow e^{-i\alpha_{ew}\gamma_{5}}q, \quad \bar{q}
\rightarrow \bar{q}e^{-i\alpha_{ew}\gamma_{5}}
\eqno\ldots(5.4)$$
There relabelling, however, introduces a Jacobian
$$J(\alpha_{ew}) = \exp\left[-i\alpha_{ew} \frac{g^{2}}{16\pi^{2}}
\epsilon_{\mu\nu\lambda\rho} \int tr F_{\mu\nu}
F_{\lambda\rho} d^{4}x\right] \eqno\ldots(5.5)$$
where the coefficient of $\alpha_{ew}$ in the exponent is $2\nu$,
i.e. twice the winding number of the gauge field configuration,
which is nontrivial precisely in sectors where instantons live.
The relabelling, therefore, merely shifts $\alpha_{ew}$ to
$\theta_{QCD}$ giving an effective $\bar{\theta}$
$$\bar{\theta} = \theta_{QCD} - 2N_{f}\alpha_{ew}
\eqno\ldots(5.6)$$
where $N_{f}$ is the number of quark flavours. All physical
quantities in this scenario, therefore, depends on $\bar{\theta}$
and not on $\theta_{QCD}$ or $\alpha_{ew}$ individually.
Theoretical estimates$^{22}$ for CP-violating EDMN are all in the
range
$$d^{th}_{n} \approx \bar{\theta} \times 10^{-15 \pm 1} e.cm
\eqno\ldots(5.7)$$
Experimental data $d_{n}^{ex} \leq 10^{-26} e.cm$, therefore,
suggests $\bar{\theta} < 10^{-9}$. Such a small value requires
near cancellation of two parameters $\theta_{QCD}$ and
$\alpha_{ew}$ as in (5.6), which arise from completely different
sectors of the Standard Model. This is the strong CP problem,
which is essentially a problem of fine tuning.

Attempts to remedy the strong CP problem by invoking a
spontaneously broken global chiral U(1) symmetry, the
Peccei-Quinn symmetry, have been pursued vigorously$^{22}$. The
idea essentially is that the effective $\bar{\theta}$ becomes a
dynamical variable in this scenario involving the field of the
pseudoscalar Goldstone boson associated with the broken
Peccei-Quinn symmetry. The dynamical $\bar{\theta}$ could then
settle down to a minimum consistent with the conservation of P
and CP. The axion has been virtually ruled out by experiments and
the strong CP problem in its original formulation is no closer to
a resolution now than it was at the time of its
conception$^{22}$.

{\bf Question of Global Chiral Anomaly} : In view of the
prevailing impasse, with axion window virtually closed, it is
worthwhile to reexamine critically the basic premises that lead
up to the strong CP problem. The question of a nontrivial global
chiral anomaly clearly stands out as the most vulnerable among
these basic premises.

The chiral limit of the axial vector Ward identity in a
instanton-like background gauge field was given in (3.23)
$$\langle\partial_{\mu}\left(\bar{\psi}\gamma_{5}\gamma_{\mu}\psi\right)\rangle_{m=0}
= 2\sum\epsilon_{i}\phi_{0i}^{+}\phi_{0i} -
\frac{g^{2}}{16\pi^{2}}\epsilon_{\mu\nu\lambda\rho}t_{r}F_{\mu\nu}F_{\lambda\rho}
\eqno\ldots(3.23)$$
where the zero modes $\phi_{0i}(x)$ are a fallout of the
nontrivial winding number $\nu$ of the gauge field. It is natural
to identify the right hand side of (3.23) as the density of
global chiral anomaly in an instanton-like background. Its
space-time integral, the global chiral anomaly, vanishes by the
Atiyah-Singer index theorem (3.19). This patently contradicts a
nontrivial Jacobian as in (5.5), the cornerstone of the strong CP
problem. The popular perception of a nontrivial global chiral
anomaly and hence a nontrivial Jacobian (5.5) not only leads to the
strong CP problem but is afflicted with contradictions in the
chiral limit.

The source of these afflictions is easily traced to the popular
identification of the partition function with the determinant of
the Dirac operator
$$Z_{f}[A]_{\nu \not= 0} = \det(i\D + m) \eqno\ldots(5.8)$$
which is unphysical in the chiral limit because of zero modes. A
key to the problem is provided by the theorem$^{23}$ which states
that there are no wrong chirality zero modes of the Dirac
operator $\D$, i.e., in the Atiyah-Singer index theorem (3.19)
positive (negative) chirality zero modes $n_{+}(n_{-})$ are
associated with positive (negative) winding number $\nu$. Thus
$$\begin{array}{ll}
\dim \ker (D_{R} D_{L}) = 0,  &\nu \geq 0 \\
\dim \ker (D_{L} D_{R}) = 0,  &\nu \leq 0 \end{array}
\eqno\ldots(5.9)$$
where $D_{L}, D_{R} = D^{+}_{L}$ are the Weyl components of the
Dirac operator $\D$
$$\D = \left(\begin{array}{cc} 0 &D_{L} \\ D_{R} &0 \end{array}\right) \eqno\ldots(5.10)$$
The theorem (5.9), therefore, assures that the partition
functions defined as
$$\begin{array}{ll}
Z_{f}[A]_{\nu \geq 0} &= \det \left(D_{R}D_{L} + m^{2}\right) \\
Z_{f}[A]_{\nu \leq 0} &= \det \left(D_{L}D_{R} + m^{2}\right)
\end{array} \eqno\ldots(5.11)$$
in the respective gauge field sectors, are not afflicted, unlike
(5.8), with zero modes and hence have smooth chiral limits. In
the trivial sector $\nu = 0$ the two representations coincide.

The representations in (5.11) require that instead of the Dirac
basis $\left\{\phi_{n}(x)\right\}$ of Sec.3 we use eigenfunction
sets of Weyl operators $D_{R}D_{L}$ and $D_{L}D_{R}$ appropriate
respectively for positive and negative $\nu$. Thus for $\nu \geq
0$, one writes
$$\phi_{n}(x) = \frac{1}{\sqrt{2}}
{\displaystyle{\frac{1}{\lambda_{n}}} 
D_{L}\phi_{nL}(x) \choose \phi_{nL}(x)}, \quad \phi_{-n}(x) =
\frac{1}{\sqrt{2}}
{\displaystyle{\frac{1}{\lambda_{n}}}D_{L}\phi_{nL}(x) 
\choose -\phi_{nL}(x)} \eqno\ldots(5.12)$$
where $\phi_{nL}(x)$ are orthonormal eigenfunctions of the positive
definite hermitian operator $D_{R}D_{L}$
$$D_{R}D_{L}\phi_{nL}(x) = \lambda^{2}_{n}\phi_{nL}(x)
\eqno\ldots(5.13)$$
The set $\{\phi_{nL}(x)\}$ with $\lambda^{2}_{n} > 0$ provides a
complete set of functions in the Weyl basis in $\nu \geq 0$
sector. In the resulting axial vector Ward identity$^{24}$
$$\langle\partial_{\mu}\left(\bar{\psi}\gamma_{5}\gamma_{\mu}\psi\right)\rangle_{\nu
> 0} = 2m\langle\bar{\psi}\gamma_{5}\psi\rangle -
\left\{\frac{g^{2}}{16\pi^{2}}\epsilon_{\mu\nu\lambda\rho}t_{r}F_{\mu\nu}F_{\lambda\rho}
- \sum\phi_{0i}^{+} \phi_{0i}\right\} \eqno\ldots(5.14)$$
the contribution from the mass term on the right hand side now
vanishes smoothly in the chiral limit. The global chiral anomaly
given by the space-time integral of the chiral limit of the four
divergence of the axial vector current, therefore, vanishes and
instead of (5.5), we now have

$$\begin{array}{ll}
J(\alpha_{ew})_{\nu\not=0} &= \displaystyle\exp\left[-i\alpha_{ew}
\int\left\{\frac{g^{2}}{16\pi^{2}}
\epsilon_{\mu\nu\lambda\rho} tr F_{\mu\nu} F_{\lambda\rho} -
\sum\epsilon_{i}\phi^{+}_{0i}\phi_{0i}\right\}d^{4}x\right] \\
&= 1 \end{array} \eqno\ldots(5.15)$$

The vanishing of the global chiral anomaly means that the chiral
phase $\alpha_{ew}$ in the quark mass in (5.1) is unphysical and
can be transformed away trivially by a global chiral rotation
(5.4) without affecting in any way $\theta_{QCD}$. The vacuum
parameter $\theta_{QCD}$ remains invariant. The crux of the strong
CP problem, the problem of fine tuning, therefore, melts away.
CP symmetry is ensured simply through the natural choice
$\theta_{QCD} = 0$.

\section{Concluding Remarks}
Ever since its conception in the context of the problem of
neutral pion decay into two photons, chiral anomaly has been a
topic of abiding interest and challenge in particle physics. The
interest stems in a large measure from the need to couple
fermions chirally to gauge fields in building models in particle
physics. The challenge consists in formulating a consistent and
systematic regularisation scheme in chiral gauge theories.

The paper highlights and elucidates the seminal role of the mass
term in the axial vector Ward identity in generating the local
ABJ anomaly and the global U(1) axial anomaly. Gauge invariance
demands that the fermion gets decoupled from the divergence of
the U(1) axial vector current if it is very heavy. This
identifies the ABJ anomaly with the asymptotic limit of the
fermion mass term with sign reversed. On the other hand, the
chiral limit $(m = 0)$ of the same mass term does not vanish and
consists of contributions from fermion zero modes when the
background gauge field has a nontrivial topology $\nu \not= 0$. The
space time integral of the chiral limit cancels the integral of
the ABJ anomaly, the (sign--reversed) asymptotic limit of the
mass term, thanks to the Atiyah-Singer index theorem. This
suggests, contrary to popular perception, that the Jacobian for
global U(1) chiral transformation is trivial even in an instanton
background. The triviality of the Jacobian is realised in a
representation of the fermion partition function in the Weyl
basis (5.11) which has a null kernel space. The point of interest
in all this is that there is no strong CP problem in an axionless
physical world.

Current interest in lattice formulation of chiral gauge theory
centres around Dirac operators for lattice fermion which obey the
Ginsparg-Wilson$^{16}$ relation (4.4). Apart from redefining
chiral symmetry on lattice, the Ginsparg-Wilson relation
introduces nonlocality$^{18}$. It is interesting to note that in
continuum formulation also it is possible to define a gauge
invariant partition function (4.8) but only at the cost of
locality. Fermion averaging of the gauge current implemented with
this partition function yields the covariant current. The
consistent current which obeys integrability, can be generated
with the covariant current as input. The covariant derivative of
the consistent current thus obtained yields the minimal anomaly
which obeys the Wess-Zumino consistency condition. Both the
anomalies, covariant and consistent, and the distinction between
the two currents vanish if the fermion belongs to anomaly free
representation (4.26).

\newpage

\centerline{\large\bf Acknowledgement}

It is a pleasure to acknowledge indebtedness to my collaborators
Rabin Banerjee, Asit De, and Partha Mitra. I should also like to
thank Amitabha Lahiri for discussions, and Sugata Mukherjee and
S.K. Singh for help in preparing the manuscript.

\end{document}